\documentclass[aps,pra]{revtex4}
\usepackage{graphicx}

\begin{document}

\title{Impact of the nuclear mass uncertainties on the $r$ process}

\author{Z. Y. Wang}
\author{Q. G. Wen}
\author{T. H. Heng}\email{hength@ahu.edu.cn}
\affiliation{School of Physics and Materials Science, Anhui
University, Hefei 230039, China}

\date{\today}
\maketitle

\begin{minipage}{420pt}

{\bf Abstract} Based on a simple site-independent approach, we
attempt to reproduce the solar $r$-process abundance with four
nuclear mass models, and investigate the impact of the nuclear mass
uncertainties on the $r$ process. In this paper, we first analyze
the reliability of an adopted empirical formula for $\beta$-decay
half-lives which is a key ingredient for the $r$ process. Then we
apply four different mass tables to study the $r$-process
nucleosynthesis together with the calculated $\beta$-decay
half-lives, and the existing $\beta$-decay data from FRDM+QRPA is
also considered for comparison. The numerical results show that the
main features of the solar $r$-process pattern and the locations of
the abundance peaks can be reproduced well via the $r$-process
simulations. Moreover, we also find that the mass uncertainties can
significantly affect the derived astrophysical conditions for the
$r$-process site, and resulting in a remarkable impact on the $r$ process. \\


\end{minipage}

\section{Introduction}

It has long been thought that the rapid neutron capture process
($r$-process) is responsible for the synthesis of half of the heavy
elements beyond iron~\cite{Burbidge1957RMP, Cameron1957CRL}. In the
$r$ process, the iron group or the nuclei up to $A\approx90$ are
regarded as seed nuclei, and then capture the neutrons in time
scales shorter than $\beta$ decay. In this way the corresponding
pathway of $r$-process involves many neutron-rich nuclei which are
far away from the $\beta$ stability, even close to the neutron
drip-line. Unfortunately only few of these very exotic nuclei can be
produced in current or even next-generation rare isotope beam
facilities. As a result, calculations based on theoretical models of
the reaction chain are essential to understand the origin of the
heaviest nuclei in the universe and the $r$-process nucleosynthesis.

As to the $r$-process nucleosynthesis, one of the underlying
difficulties is the astrophysical sites, which have not yet been
unambiguously identified up to now. However, there is common
agreement that the $r$-process occurs on the premise of extreme
neutron densities, and then runs through very neutron-rich nuclei
far-off the valley of stability~\cite{Cowan1991PR}. So far many
studies have been performed for the candidate
sites~\cite{Meyer1992AJ, Takahashi1994AA, Qian1996AJ,
Freiburghaus1999AJ, Goriely2005NPA, Sumiyoshi2001AJ, Wanajo2003AJ,
MacFadyen1999AJ, Pruet2004AJ, Ning2007AJL, Hu^{"}depohl2010PRL,
Caballero2012AJ} (for examples, the "neutron star (NS) merger"
~\cite{Freiburghaus1999AJ, Goriely2005NPA}, the "prompt explosion"
from a low mass SN~\cite{Sumiyoshi2001AJ, Wanajo2003AJ}, and the
"collapsar" from a massive progenitor~\cite{MacFadyen1999AJ,
Pruet2004AJ}). Nevertheless each of them still faces severe problems
and cannot be asked to explain the production of the $r$-process
nuclei observed in nature, and leading to a lack of consensus at the
present time.

The other puzzle is the nondeterminacy of the nuclear properties
which are essential for the $r$-process and can not be gained using
the current experimental techniques. In particular, mass predictions
for neutron-rich nuclei have a vital influence on the relevant
nuclear reactions in $r$-process, such as the photodisintegration,
the neutron-capture, the fission probabilities, and the
$\beta$-decay rates. In view of the importance of nuclear mass in
the $r$ process, many related theoretical calculations were
performed in the past decades. One conventional method is the local
mass relations, which have a high precision of prediction for nearby
nuclei, such as the Garvey-Kelson relations~\cite{Garvey1969RMP},
residual proton-neutron interactions~\cite{Zhang1989PLB, Fu2010PRC},
Coulomb-energy displacement~\cite{Sun2011SCPMA, Kaneko2013PRL}, and
systematics of $\alpha$-decay energies~\cite{Dong2011PRL}. The other
method depends on the global mass models. These mass models are
generally regarded to have a better extrapolation for nuclei far
from the known region, for examples, the macroscopic-microscopic
finite-range droplet model (FRDM)~\cite{Mo^{"}ller1995ADNDT}, the
Weizs\"{a}cker-Skyrme (WS) model~\cite{Wang2014PLB}, the microscopic
Hartree-Fock-Bogoliubov (HFB) theory with a Skyrme
force~\cite{Goriely2013PRC} and the relativistic mean-field (RMF)
model~\cite{Geng2005PTP, Hua2012SCPMA, Arteaga2016EPJA}. Compared to
the former mass relations, the accuracies of these global mass
models are worse. Therefore, several methods have been introduced to
improve their accuracies, such as the CLEAN image reconstruction
technique~\cite{Morales2010PRC}, the radial basis function
approach~\cite{Wang2011PRC, Niu2013PRCb, Zheng2014PRC, Niu2016PRC,
Niu2018SciB}, and the neural network approach~\cite{Utama2016PRC,
Zhang2017JPG, Niu2018PLB}.

After Burbidge's systematic introduction to the $r$ process for the
first time~\cite{Burbidge1957RPM}, due to the lag of experimental
and theoretical development, only the phenomenological nuclear
droplet mass formula~\cite{Hilf1976CERN} could be considered for the
$r$-process calculations in a long period. Fortunately by now, the
theoretical study of nuclear properties has made a great progress
which results in lots of various $r$-process
calculations~\cite{Pfeiffer1997ZPA, Wanajo2004AJ, Sun2008PRC,
Niu2009PRC, Xu2013PRC}. In addition to the mass, the $\beta$-decay
half-life plays key role in estimating the $r$-process abundances,
which determines the time-scale for the matter flow from the seeds
to the heavy nuclei. In fact, the evaluation of $\beta$-decay rates
for the waiting-point nuclei is one of the important issues of the
nucleosynthesis through the $r$ process. As we known, the famous
Fermi theory~\cite{Fermi1934ZP} in the 1930's is commonly regarded
as the starting point of the theoretical investigations for the
nuclear $\beta$-decay. Nowadays, as the extremes, there are mainly
two types of microscopic approaches for the large-scale calculations
of nuclear $\beta$-decay half-lives: the shell model and the
proton-neutron quasiparticle random-phase approximation (QRPA). Due
to extremely large configuration spaces, the shell-model
calculations can not be applied to the heavy nuclei far from the
magic numbers. In contrast, the proton-neutron QRPA are feasible for
the arbitrary heavy systems~\cite{Engel1999PRC, Liang2008PRL,
Niu2017PRC}.

Thus far, the nuclear $\beta$-decay calculations have already been
carried out using the QRPA based on the
FRDM~\cite{Moller1997ADNDTMoller2003PRC}, the extended Thomas¨CFermi
plus Strutinsky integral (ETFSI) model~\cite{Borzov2000PRC}, the
Skyrme Hartree¨CFock¨CBogoliubov (SHFB) model~\cite{Engel1999PRC},
the density functional of Fayans (DF)~\cite{Borzov1996ZPA}, and the
covariant density functionals~\cite{Niu2013PLB, Niu2013PRCR,
Marketin2016PRC}. Recently, Zhang $\emph{et
al.}$~\cite{Zhang2006PRCZhang2007JPGNPP} presented a new exponential
law for calculating $\beta$-decay half-lives of nuclei far from the
stability line. In 2017, Zhou $\emph{et al.}$~\cite{Zhou2017SCPMA}
proposed an empirical formula for $\beta$-decay half-lives via
investigating systematically the variation of $\beta$-decay
half-lives with the decay energy $Q$ and nucleon number $(Z, N)$
based on experimental data. Here both shell and pairing effects on
$\beta$-decay half-lives versus the nucleon number $(Z, N)$ are
taken into account.

In this paper, based on Zhou's empirical formula for $\beta$-decay
half-lives, we consider four nuclear mass models to provide the
theoretical $Q$-values for $\beta$-decay half-lives. In order to
testify the precision of this formula for reliably predicting
$\beta$-decay half-lives in this paper, comparisons about the
root-mean-square (rms) deviation between the calculated
$\beta$-decay half-lives and the experimental
data~\cite{Audi2017CPC}, together with the existing $\beta$-decay
properties from FRDM+QRPA~\cite{Moller1997ADNDTMoller2003PRC}, are
performed. Then with the calculated $\beta$-decay half-lives, four
nuclear mass models are applied to investigate the impact of
theoretical uncertainty of unknown masses in the $r$-process
calculation. The numerical results show that the acquired
astrophysical conditions for the $r$-process are significantly
different with each other and very sensitive to the adopted mass
models, and the impact of mass uncertainties on the solar
$r$-process abundances is very important to completely understand
the $r$-process.

This paper is organized as follows. In section II, the calculating
formula for $\beta$-decay half-lives and the introduction to a
site-independent $r$-process approach are briefly described. In
section III, four nuclear mass models are considered to provide the
$Q$-values for $\beta$-decay half-lives with the empirical formula.
Then they are subsequently applied to reproduce the solar
$r$-process abundances. In addition, the numerical results of the
simulation results as well as some detailed discussions are given,
including some comparisons between different mass models with
(without) the calculated $\beta$-decay half-lives. Finally, a
concise summary is delivered in section IV.

\section{The theoretical method}

\subsection{The employed formula for the $\beta$ decay half-lives}

Recently, with corrections to both pairing and shell effects, Zhou
$\emph{et al.}$~\cite{Zhou2017SCPMA} proposed a formula to calculate
the $\beta^-$-decay half-lives of the neutron-rich nuclei far from
the stability line,
\begin{eqnarray}
\ln t_{1/2}&=&a_6+(\alpha^2Z^2-5-a_7\frac{N-Z}{A})\ln(Q-a_8\delta)
+a_9\alpha^2Z^2+\frac{1}{3}\alpha^2Z^2\ln(A)-\alpha Z\pi +S(Z,N),\nonumber\\
\end{eqnarray}
in which the relationship among $\beta$-decay half-lives, $Q$-values
and nucleon numbers $(Z, N)$ is detailed shown. In this equation,
$\alpha$ is the fine structure constant with value $\frac{1}{137}$.
The most important term
$(\alpha^2Z^2-5-a_7\frac{N-Z}{A})\ln(Q-a_8\delta)$ has a dominant
effect. The accuracy of $Q_\beta$-values is an strong influence on
this formula, and $Q_\beta$-values can be calculated using mass data
as $Q_\beta=M_p-M_d$, $M_p$ and $M_d$ are the masses of parent and
daughter nuclei, respectively. $\delta$ denotes the even-odd
staggering caused by pairing effects on $\beta^-$-decay half-lives
versus $Q$-values, and can be described by
\begin{eqnarray}
\delta=(-1)^N+(-1)^Z.\nonumber
\end{eqnarray}
It is pointed out that since the mass table~\cite{Wang2014PLB}
already contains the pairing energy, the pairing effects on
$Q$-values versus neutron number have already been embodied in the
calculation of $Q-$values.

As for the shell correction $S(Z, N)$, its contributions only appear
near the nucleon magic-numbers, and can be written as
\begin{eqnarray}
S(Z, N)&=&a_1e^{-((Z-20)^2+(N-28)^2)/12}\nonumber\\
&+&a_2e^{-((Z-38)^2+(N-50)^2)/43}\nonumber\\
&+&a_3e^{-((Z-50)^2+(N-82)^2)/13}\nonumber\\
&+&a_4e^{-((Z-58)^2+(N-82)^2)/24}\nonumber\\
&+&a_5e^{-((Z-70)^2+(N-110)^2)/244}.\nonumber
\end{eqnarray}

In the equation (1), all the nine parameters in Eq.(1) can be
obtained explicitly through a least-squares fit to the available
experimental data of $\beta^-$-decay half-lives~\cite{Audi2017CPC}.
In addition, as we known, the ratio of calculated value
$T^{\textrm{theo.}}_{1/2}$ to experimental value
$T^{\textrm{expt.}}_{1/2}$ of $\beta$-decay half-lives can reflect
the ability of reproducing experiment data and the extrapolation
capacity via this analytical formula. So we defined the rms
deviation of the decimal logarithm of the $\beta$-decay half-life,
in this paper, as
\begin{eqnarray}
\sigma=\sqrt{\frac{1}{N}\sum^{N}_{i=1}[\log_{10}(T^{\textrm{theo.}}_{1/2
i})-\log_{10}(T^{\textrm{expt.}}_{1/2 i})]^2}.
\end{eqnarray}
$N$ is the number of nuclei used for evaluation of the rms
deviation. The acquired experimental $\beta^-$-decay half-lives are
taken from Ref.~\cite{Audi2017CPC}.

\subsection{The site-independent $r$-process approach}

The $r$ process not only depends on various inputs of nuclear
properties, but is also affected by multifarious astrophysical
conditions: the entropy and temperature of the explosion
environment, the electron-to-baryon number ratio ($Y_e$), and the
neutrino processes, etc. It is unsatisfied that the astrophysical
sites for $r$-process nucleosynthesis have still not been
unambiguously stated. Even many candidate sites have been tried to
proposed and supernovae appears to be well suited as the $r$-process
site~\cite{Mathews1990N}. However, up to now there has been no
conclusion as the correct astrophysical model. In this paper, we
employed a site-independent approach (for recent reviews, see, e.g.,
Refs.~\cite{Kratz2007AJ, Sun2008PRC, Cowan1999AJ, Pfeiffer1997ZPA}),
in which the solar $r$-process abundances~\cite{Simmerer2004AJ} have
been used to constrain the astrophysical conditions, to calculate
the $r$-process abundance with different nuclear physics inputs

In this approach, it is supposed that neutron sources irradiate the
seed nuclei over a time scale $\tau$ in a high-temperature
environment (T $\sim$ 1 GK). It is particularly important that the
neutron sources have high and continuous neutron densities $n_n$
ranging from $10^{20}$ to $10^{28}$ $cm^{-3}$. Because of the very
high neutron densities, the $\beta$ decays will be largely
overwhelmed by the competing neutron captures, leading to the
equilibrium $(n, \gamma)\Longleftrightarrow (\gamma,n)$ for every
element. The abundance ratio of two isotopes on a time scale $\tau$
can be written as
\begin{eqnarray}
\frac{Y(Z,A+1)}{Y(Z,A)}&=&n_n(\frac{h^2}{2\pi m_\mu\kappa
T})^{3/2}\frac{G(Z,A+1)}{2G(Z,A)}(\frac{A+1}{A})^{3/2}
\times\exp[\frac{S_n(Z,A+1)}{\kappa T}].
\end{eqnarray}
In the above equation, the symbols $Y(Z, A)$, $S_n$, $G(Z, A)$
denotes the abundance of the nuclide $(Z, A)$, the one-neutron
separation energy and the partition function of nuclide $(Z, A)$,
and $h$, $\kappa$, and $m_\mu$ are the Planck constant, Boltzmann
constant, and atomic mass unit,respectively. In addition, suppose
$Y(Z, A + 1)/Y(Z, A)$ is close to $1$ at the highest isotopic
abundance for each element, and all other quantities are fixed to be
constant, the average neutron-separation energy ${\overline S}_n$
will be the same for all the nuclides with the highest abundance for
each isotopic chain. In this method, ${\overline S}_n$ is defined as
\begin{eqnarray}
{\overline S}_n&\approx& \kappa T \log[\frac{2}{n_n}(\frac{2\pi
m_\mu\kappa T}{h^2})^{3/2}]\nonumber\\
&=&T_9\{2.79+0.198[\log(\frac{10^{20}}{n_n})+\frac{3}{2}\log(T_9)]\},
\end{eqnarray}
where $T_9$ correspond to the temperature in $10^9$ K.

While we ignore the fission reaction, the matter flow in $r$ process
is governed by $\beta$ decays. The corresponding abundance can be
shown using a group of differential equations:
\begin{eqnarray}
\frac{dY(Z,A)}{dt}&=&Y(Z-1)\sum_{A}P(Z-1,A)\lambda^{Z-1,A}_{\beta}
-Y(Z)\sum_{A}P(Z,A)\lambda^{Z,A}_{\beta},
\end{eqnarray}
here $Y(Z)=\sum_{A}Y(Z, A)=\sum_{A}P(Z, A)Y(Z)$ is the total
abundance for each isotopic chain ($P(Z, A)$ is the isotopic
abundance distribution). $\lambda^{Z,A}_{\beta}$ is the total decay
rate of the nuclide $(Z, A)$ via the delayed neutron emission or the
$\beta$ decay. Then after the neutrons freeze out, all the isotopes
will go to the corresponding stable elementary substance via the
$\beta$ decays, for which the abundance can be obtained based on the
two equations (3) and (5).

\section{Results and discussions}

\subsection{The reliability of the systematic formula}

\begin{table}
\begin{center}
\caption{Comparisons between the rms deviations of the experimental
$\beta^-$-decay half-lives~\cite{Audi2017CPC} and the calculated
data from four nuclear mass tables, together with the deviations
arising from FRDM+QRPA and the Eq.(1) itself. The first column
includes the symbols for the corresponding nuclear mass tables, the
extractive $\beta$-decay properties from
Ref.~\cite{Moller1997ADNDTMoller2003PRC} (FRDM+QRPA) and the
equation 1 itself (EQ.(1)). The first row contains the adopted
regions of experimental $\beta$-decay half-life values. See text for
more details.} \label{ttb1}
\begin{tabular}{lccccccccccccccc}
\hline\hline & $T\leq 1$ s  & $T\leq 1000$ s  \\ \hline
FRDM   & 0.381 & 0.598    \\
RMF  & 0.433 & 0.730  \\
HFB-31 & 0.375 & 0.651 \\
WS4 & 0.316 & 0.529 &  \\
Eq.(1) & 0.364 & 0.584 &  \\
FRDM+QRPA & 0.391 & 0.597  \\
\hline\hline
\end{tabular}
\end{center}
\end{table}

In the $r$ process, there are two reasons why $\beta$-decay
lifetimes are regarded as the critical inputs. The first one is that
they set the timescale for heavy element production if the
equilibrium $(n, \gamma)\Leftrightarrow(\gamma, n)$ occurs. Another
is that $\beta$-decay lifetimes help to shape the final pattern as
the path moves back to stability. In this paper, the needed
$\beta$-decay lifetimes will be calculated by using the empirical
formula~\cite{Zhou2017SCPMA}. However, it is doubtful to what extent
the predictive ability of the adopted formula in this paper can
achieve. According to the formula expressed in Eq.(1), one can see
half-life calculations depend somewhat on the values of the
parameter $Q$, then reliable theoretical predictions of the nuclear
mass, bringing about precise $Q$-values, are essential to study
$\beta$ transitions when these nuclear masses can not be given
experimentally.

In this subsection, four different mass tables, i.e., the FRDM
model~\cite{Mo^{"}ller1995ADNDT}, the latest version of WS (WS4)
model~\cite{Wang2014PLB}, the recent version of the HFB model
(HFB-31)~\cite{Goriely2016PRC}, and the RMF model with TMA effective
interaction~\cite{Geng2005PTP}, are considered to demonstrate the
predictive ability of the systematic formula~\cite{Zhou2017SCPMA}.
The model deviation of binding energy $B$ with respect to the
experimental data can be characterized by the rms deviation
$\sigma_{rms}=\sqrt{\frac{1}{N}\sum^{N}_{i=1}(B^{\textrm{theo.}}-B^{\textrm{expt.}})_i^2}$,
and the corresponding numerical numbers for the FRDM, WS4, HFB-13,
and RMF with respect to experimentally determined
values~\cite{Wang2017CPC} are $0.677$, $0.302$, $0.584$ and $2.258$
MeV, respectively. The other model deviations of
$\log_{10}(T^{\textrm{theo.}}_{1/2}/T^{\textrm{expt.}}_{1/2})$ for
these four mass models and the Eq.(1) itself are list in
Tab.~\ref{ttb1}, and for comparison we also take into account the
rms deviations between the existing $\beta$-decay properties taken
from Ref.~\cite{Moller1997ADNDTMoller2003PRC} (FRDM+QRPA) and recent
experiment data~\cite{Audi2017CPC}. By the way, according to the
updated information of Audi $\emph{et al.}$~\cite{Audi2017CPC},
there are totally about 1101 nuclei for all possible $\beta^{-}$
transitions with well-defined half-lives ($T$). The eligible data
set adopted in this paper contains 824 nuclei with their $Z$, $N$
and half-life values being in the selected regions of $Z\geq 8$,
$N\geq 8$ and $T\leq 1000 $ second (s), respectively. Furthermore,
for the sake of illustration, two categories of comparisons are
performed according to the regions of adopted experimental
$\beta^-$-decay half-life values $T$, i.e., $T\leq 1 (s)$ and $T\leq
1000 (s)$.

As can be seen in Tab.~\ref{ttb1}, in the first category of $T\leq
1$ s (including 381 nuclei), the WS4
model~\cite{Wang2014PLB},combined with the employed
formula~\cite{Zhou2017SCPMA}, gives the smallest rms values of
$0.316$, which means the overall ratios between calculated
$\beta^-$-decay half-life values and experimental ones are about a
little more than twice. For FRDM~\cite{Mo^{"}ller1995ADNDT},
HFB-31~\cite{Goriely2016PRC} and RMF~\cite{Geng2005PTP}, the
corresponding rms deviations are $0.381$, $0.375$ and $0.433$,
respectively, less than the rms deviation of $0.391$ deduced from
traditional FRDM+QRPA~\cite{Moller1997ADNDTMoller2003PRC} except for
RMF~\cite{Geng2005PTP}. The rms deviation obtained using the
empirical formula~\cite{Zhou2017SCPMA} and the experimental
$Q$-values~\cite{Wang2017CPC} is $0.364$, also lower than that of
FRDM+QRPA~\cite{Moller1997ADNDTMoller2003PRC}. All of these mean
that in this category of $T\leq 1$ s, the best estimations of
$\beta^-$-decay half-lives come from WS4~\cite{Wang2014PLB} while
the worst one from RMF~\cite{Geng2005PTP}. For
FRDM~\cite{Mo^{"}ller1995ADNDT}, HFB-31 and the Eq.(1) itself,
anyone of them has a higher precision of prediction for
$\beta^-$-decay half-lives than
FRDM+QRPA~\cite{Moller1997ADNDTMoller2003PRC}. In another category
of $T\leq 1000$ s (including 824 nuclei), one can see that
WS4~\cite{Wang2014PLB} still provides the optimal outcome with the
$\sigma$-values being only $0.529$. The following one is $0.584$
deduced from the Eq.(1) itself~\cite{Zhou2017SCPMA} and the
experimental $Q$-values~\cite{Wang2017CPC}.
FRDM+QRPA~\cite{Moller1997ADNDTMoller2003PRC} wins third place with
the corresponding value being $0.597$. While considering
FRDM~\cite{Mo^{"}ller1995ADNDT}, HFB-31~\cite{Goriely2016PRC} and
RMF~\cite{Geng2005PTP}, the calculated rms deviations are $0.598$,
$0.651$ and $0.730$, respectively, all higher than that of
FRDM+QRPA~\cite{Moller1997ADNDTMoller2003PRC}. These mean that the
similar conclusions for both WS4 model~\cite{Wang2014PLB} and the
Eq.(1) itself~\cite{Zhou2017SCPMA} can be attained in comparison
with FRDM+QRPA~\cite{Moller1997ADNDTMoller2003PRC}. But for
FRDM~\cite{Mo^{"}ller1995ADNDT} and HFB-31~\cite{Goriely2016PRC},
the opposite is true. That is with the empirical
formula~\cite{Zhou2017SCPMA}, WS4~\cite{Wang2014PLB} brings the best
accuracy of $\beta^-$-decay half-life prediction, followed by the
combination of the formula itself~\cite{Zhou2017SCPMA} and the
experimental $Q$-values~\cite{Wang2017CPC}. But the accuracies
resulted from both FRDM~\cite{Mo^{"}ller1995ADNDT} and
HFB-31~\cite{Goriely2016PRC} are slightly inferior than that from
FRDM+QRPA~\cite{Moller1997ADNDTMoller2003PRC}. As is shown,
RMF~\cite{Geng2005PTP} is still last.

Based on above, one can see that the numerical results of the rms
deviation deduced from the mass models or the empirical formula
itself proposed in Eq.(1) are very different with each other in the
two categories. However, it's worth noting that these rms values of
calculated $\beta$-decay half-lives deviating from experiment
data~\cite{Audi2017CPC} are generally within the same order of
magnitude as that resulted from
FRDM+QRPA~\cite{Moller1997ADNDTMoller2003PRC}. That is to say,
compared with the existing $\beta$-decay
properties~\cite{Moller1997ADNDTMoller2003PRC}, our evaluations of
$\beta^-$-decay half-lives using the employed formula and the four
mass models agree well with the experiment values~\cite{Audi2017CPC}
to some extent. So it is believable that the adopted systematic
formula, combining with the mass models, for reliably predicting
$\beta$-decay half-lives is effective, and the corresponding
precisions in this work are appropriate.

\begin{figure}[h]
  \includegraphics[scale=0.42]{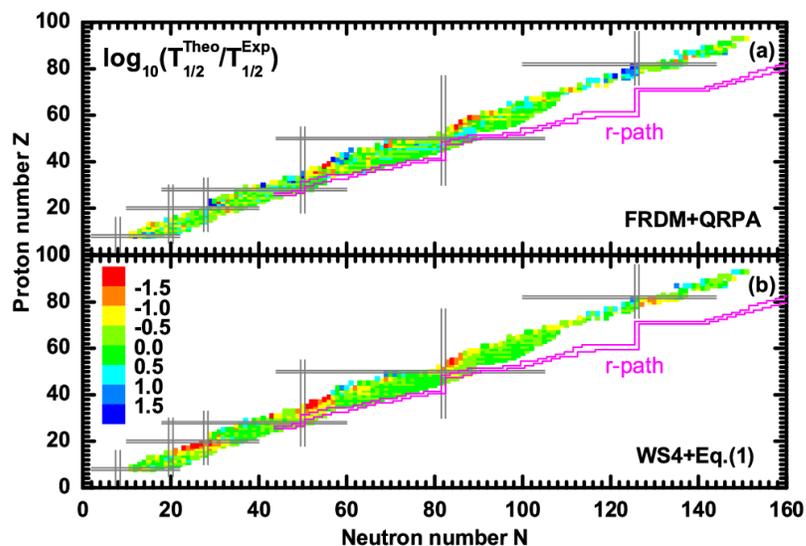}\\
  \caption{ (Color online) Decimal logarithms of $T^{theo.}_{
1/2} /T^{expt.}_{ 1/2}$ for 824 nuclei derived from WS4 model,
FRDM+QRPA and the experimental data. Panel (a) corresponds to the
results for $\beta$-decay half-lives taken from FRDM+QRPA, panel (b)
denotes the case for the calculated $\beta$-decay half-lives using
the WS4 model and the systematic formula proposed in Eq. (1),
respectively. The pairs of gray solid parallel lines denote the
magic numbers.}\label{fig1}
\end{figure}

For further insight of the extrapolating capacity of the systematic
formula, we calculated the $\beta$-decay half-lives for whole 824
nuclei, and here the WS4 model is taken as an example. Comparisons
of logarithms of the ratios between the experimental half-life
values~\cite{Audi2017CPC}, the calculated ones and the existing
$\beta$-decay properties from
FRDM+QRPA~\cite{Moller1997ADNDTMoller2003PRC} are performed,
respectively, and are drawn in Fig.1. It can be seen that the
calculated $\beta$-decay half-lives show a better agreement with the
experimental ones except the regions near magic numbers. The reason
may be that, as proposed in Ref.~\cite{Zhou2017SCPMA}, for
extrapolation to nuclei far from $\beta$-stability line, enormous
differences in tendency are evident between the results with this
formula and those from the exponential
law~\cite{Zhang2006PRCZhang2007JPGNPP}. As a result, The correction
of both the shell and pairing effects on $\beta$-decay half-lives
for the nuclei near the magic numbers in this work is not perfect.
However, compared to the case of FRDM+QRPA, the effect of systematic
correction is notable, all of which provide strong support for the
reliability of the systematic formula and its usefulness for
eliminating the discrepancy.

\subsection{The influence of mass uncertainties on the $r$-process}

As we known, the evaluation of mass formulae and $\beta$-decay
rates, particularly at the waiting point nuclei, is one of the
important issues of the nucleosynthesis through the $r$ process. In
this section, four nuclear mass tables proposed above will be
considered to reproduce the solar $r$-process abundance. The
required $\beta$-decay half-lives will be obtained using the
empirical formula and the $Q$-values from the mass tables themselves
and, to have a comparison, the existing $\beta$-decay properties
from FRDM+QRPA are considered as well.

Similar to the method used in Refs.~\cite{Kratz2007AJ, Sun2008PRC,
Cowan1999AJ, Pfeiffer1997ZPA}, 16 components with neutron densities
in the range of $10^{20}$ to $3\times 10^{27}$ $cm^{-3}$ are applied
in our calculations. Then a temperature $\emph{T}=1.5$ GK is chosen,
for which we suppose the corresponding weight $\omega$ and the
irradiation time $\tau$ follow an exponential dependence on neutron
density $n_n$,
\begin{eqnarray}
\omega(n_n)=d\times n^a_n, \hskip 0.5cm  \tau(n_n)=b\times n^c_n,
\end{eqnarray}
the four parameters $a$, $b$, $c$ and $d$ can be obtained by fitting
the solar $r$-process abundances with a least-square fit.
Furthermore, it is also assumed that the longest neutron irradiation
time should be longer than $0.5$ s but shorter than $20$ s.

\begin{figure}[h]
  \includegraphics[scale=0.42]{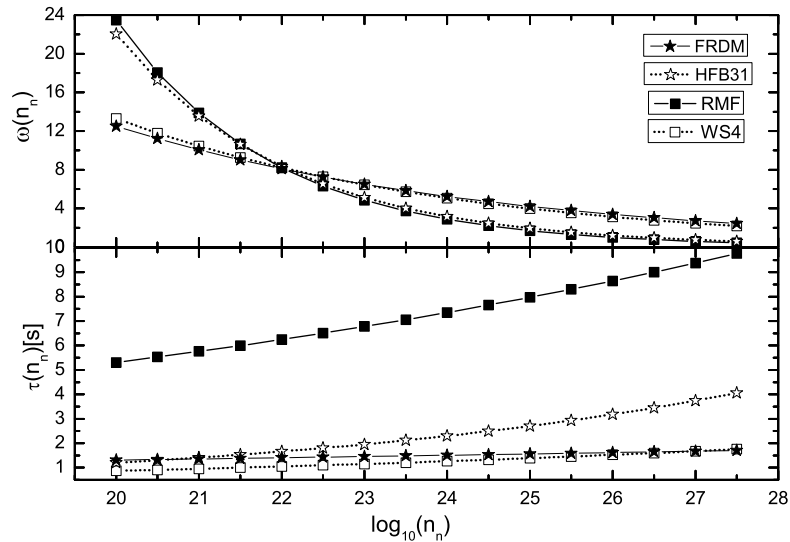}\\
  \caption{ The configuration of sixteen $r$-process components
that recur the r-process abundances with different mass inputs. The
neutron density $n_n$ is in units of $cm^{-3}$. As a function of the
neutron density $n_n$, we display the weighting factor $\omega(n_n)$
and the neutron irradiation time $\tau(n_n)$ in the upper and lower
panel, respectively. The total weighting factor has been normalized
to 100.}\label{fig2}
\end{figure}

In order to investigate the impact of theoretical uncertainties of
unknown masses in the $r$-process calculations, we take our best
simulation using the four mass models, together with the calculated
$\beta$-decay half-lives. The astrophysical conditions resulted from
the various mass inputs are shown in Fig. 2. As it is shown, in the
upper panel, the weighting factors for FRDM are almost completely
overlaid by those for WS4, and so are RMF and HFB31 models.
Moreover, the astrophysical conditions obtained from the former two
mass-model simulations require a smaller weighting factor for low
neutron density than those from the latter ones, while the case is
reverse for high neutron density. On the other hand, in the lower
panel, the astrophysical conditions found using the FRDM and WS4
mass inputs favor a relatively constant neutron irradiation time for
different neutron densities, and for the HFB-31 case, the time
becomes longer, up to $4$ s, as the neutron density increases. The
longest neutron irradiation time comes from the simulation using the
RMF model, which requires component durations of as long as $10$ s.
As a result, it is clear that the acquired astrophysical conditions
for the $r$-process are significantly different with each other and
very sensitive to the adopted mass models.

\begin{figure}[h]
  \includegraphics[scale=0.38]{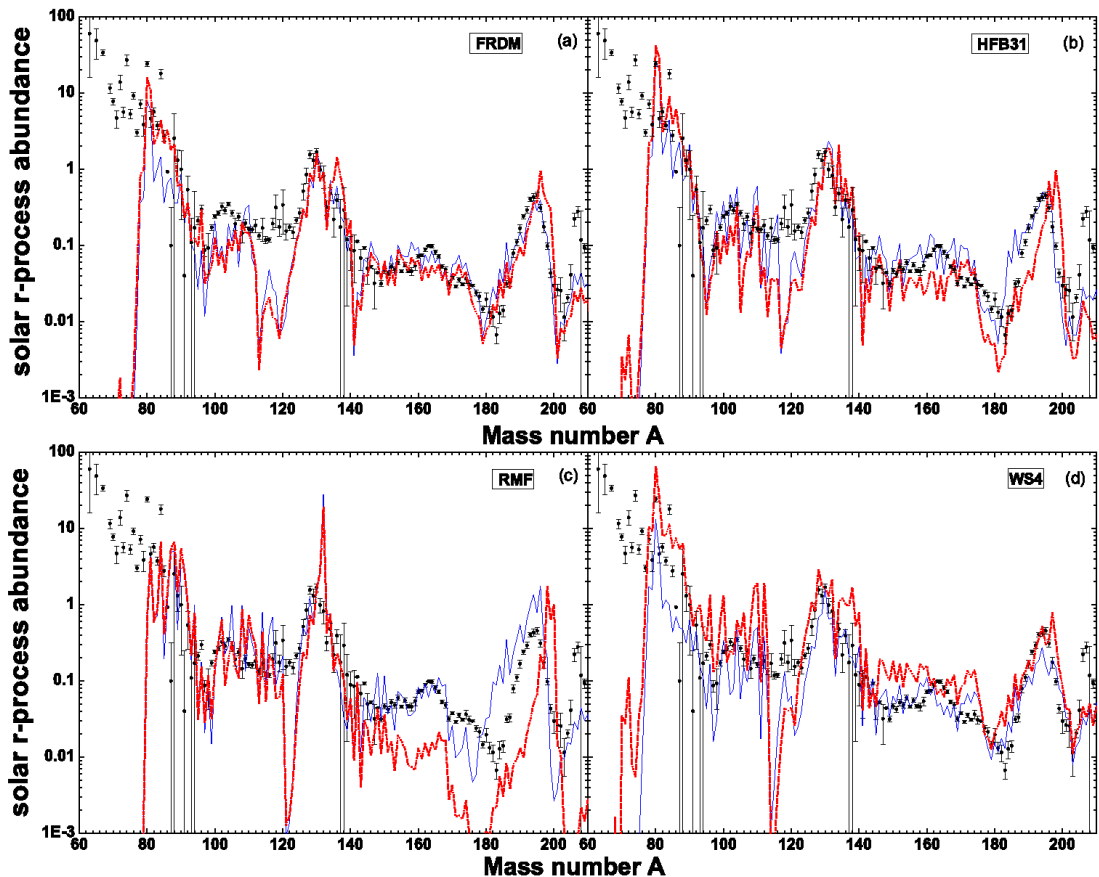}\\
  \caption{ (Color online)Simulation results of the solar $r$-process abundances (in the
logarithm scale) using different nuclear mass inputs. The needed
$\beta$-decay properties include two cases: and the existing
$\beta$-decay half-lives from FRDM+QRPA and the calculated ones with
the empirical formula plus the $Q$-values from four nuclear mass
tables. The blue-solid lines display our fits to the solar
$r$-process abundances with the calculated $\beta$-decay half-lives
while the red-dashed lines correspond to the results with another
$\beta$-decay data. The black dots with a certain range of
uncertainties designates the experimental data of solar $r$-process
abundances~\cite{Simmerer2004AJ}. See text for more details.
}\label{fig3}
\end{figure}

The solar $r$-process abundances calculated by using different mass
models and the calculated $\beta$-decay half-lives are displayed in
Fig. 3, and the results obtained with the existing $\beta$-decay
properties from FRDM+QRPA are also plotted for comparison. From Fig.
3, one can see, the results of the $r$-process abundance
calculations with various nuclear mass models differ from each
other, however, all of them bring about an abundance underproduction
at $A\sim120$, which has traditionally been put down to the
overestimated strength of the $N=82$ shell
closure~\cite{Pfeiffer1997ZPA, Cowan1999AJ} in the theoretical
nuclear physics model. Compared to the results with the existing
$\beta$-decay properties from FRDM+QRPA, the $r$-process simulations
with our calculated $\beta$-decay half-lives agree better with the
observation for all nuclear mass models, particularly for RMF, the
results with the existing $\beta$-decay properties from FRDM+QRPA
lead to a significant underestimation of the isotopic abundances
between the $A\sim130$ peak and the $A\sim195$ peak, then with our
calculated $\beta$-decay have-lives, the abundance trough is largely
filled in, as shown in Fig. 3(c). Moreover, it also happened for
HFB-31, but to a lesser degree (see Fig. 3(b)). As to nuclear mass
input, one can see, the best agreement with the solar abundance
pattern comes from WS4 models, followed by FRDM and HFB-31, the
worst one is produced by RMF, particularly between the $A\sim130$
peak and the $A\sim195$ peak.

\section{Summary}

In summary, we first adopt an empirical formula proposed by Zhou
$\emph{et al.}$~\cite{Zhou2017SCPMA} for calculating $\beta$-decay
half-lives with four nuclear mass tables as well as the experimental
$Q$-values~\cite{Wang2017CPC}. Then we compare the calculated
results with the experimental data, together with the existing
$\beta$-decay properties from FRDM+QRPA, to demonstrate the
precision of the formula for reliably predicting $\beta$-decay
half-lives. It is shown that the adopted formula is reliable and
useful for eliminating the discrepancy.

Subsequently, in order to investigate the impact of theoretical
uncertainties of unknown masses in $r$-process calculations, we
apply four nuclear mass models to reproduce the main features of the
solar $r$-process pattern and the locations of the abundance peaks
based on the site-independent $r$-process approach. The required
$\beta$-decay half-lives are calculated using the empirical formula
with the theoretic $Q$-values come from these mass tables and, for
the sake of comparison, also include the existing $\beta$-decay
properties taken from FRDM+QRPA. As is shown above, compared with
the results using the existing $\beta$-decay properties, the
$r$-process simulations with the calculated data can lead to a
better agreement with the observation for all four mass tables,
particularly for RMF, the significant underestimation of the
isotopic abundances can be largely corrected between the $A\sim130$
peak and the $A\sim195$ peak. Furthermore, we have also compared the
results of $r$-process simulations obtained from different mass
models. One can see that the deduced astrophysical conditions for
the $r$-process are significantly different and change along with
the used mass model. The $r$-process abundances obtained from WS4
agree best with the observed data, followed by FRDM and HFB-31, the
worst one is produced by RMF, all of which provide strong support
for a remarkable impact of mass uncertainties on the $r$-process
nucleosynthesis.\\

\section{Acknowledgements}

This work was supported by the National Natural Science Foundation
of China (Grant No.11505001), and the Key Research Foundation of
Education Ministry of Anhui Province of China (Grant No.
KJ2015A041).




\begin{thebibliography}{99}

\bibitem{Burbidge1957RMP} E. M. Burbidge, G. R. Burbidge, W. A. Fowler and F. Hoyle, \emph{Rev.
Mod. Phys.} \textbf{29} (1957) 547,
doi:https://doi.org/10.1103/RevModPhys.29.547.
\bibitem{Cameron1957CRL} A. G. W. Cameron, Chalk River Report \textbf{CRL}-41, 1957.
\bibitem{Cowan1991PR} J. J. Cowan, F. -K. Thieleman and J. W. Truran, \emph{Phys. Rep.} \textbf{208}
(1991) 267, doi:https://doi.org/10.1016/0370-1573(91)90070-3.
\bibitem{Meyer1992AJ}  B. S. Meyer, G. J. Mathews, W. M. Howard, S. E. Woosley and R.
D. Hoffman, \emph{Astrophys. J.} \textbf{399} (1992) 656,
doi:10.1086/171957.
\bibitem{Takahashi1994AA} K. Takahashi, J. Witti and H.-Th. Janka, \emph{Astron. Astrophys.} \textbf{286} (1994) 857,
\bibitem{Qian1996AJ} Y.-Z. Qian and S. E. Woosley, \emph{Astrophys. J.} \textbf{471}
(1996) 331, doi:http://dx.doi.org/10.1086/177973.
\bibitem{Freiburghaus1999AJ} C. Freiburghaus, S. Rosswog and F. -K. Thielemann, \emph{Astrophys.
J.} \textbf{525} (1999) L121, doi:http://dx.doi.org/10.1086/312343.
\bibitem{Goriely2005NPA} S. Goriely, P. Demetriou, H. -T. Janka, J. M. Pearson and M.
Samyn, \emph{Nucl. Phys. A} \textbf{758} (2005) 587,
doi:10.1016/j.nuclphysa.2005.05.107.
\bibitem{Sumiyoshi2001AJ} K. Sumiyoshi, M. Terasawa, G. J. Mathews, T. Kajino, S. Yamada
and H. Suzuki, \emph{Astrophys. J.} \textbf{562} (2001) 880,
doi:http://dx.doi.org/10.1086/323524.
\bibitem{Wanajo2003AJ} S. Wanajo, M. Tamamura, N. Itoh, K. Nomoto, I. Ishimaru, T. C.
Beers and S. Nozawa, \emph{Astrophys. J.} \textbf{593} (2003) 968,
doi:http://dx.doi.org/10.1086/376617.
\bibitem{MacFadyen1999AJ}A. I. MacFadyen and S. E. Woosley,\emph{ Astrophys. J.} \textbf{524}
(1999) 262, doi:http://dx.doi.org/10.1086/307790.
\bibitem{Pruet2004AJ} J. Pruet, T. A. Thompson and R. D. Hoffman, \emph{Astrophys. J.}
\textbf{606} (2004) 1006, doi:http://dx.doi.org/10.1086/382036.
\bibitem{Ning2007AJL} H. Ning, Y. -Z. Qian and B. S. Meyer, \emph{Astrophys. J. Lett.}
\textbf{667} L159 (2007), doi:http://dx.doi.org/10.1086/522372.
\bibitem{Hu^{"}depohl2010PRL} L. H\"{u}depohl, B. M\"{u}ller, H. Janka, A. Marek and G. G.
Raffelt, \emph{Phys. Rev. Lett.} \textbf{104} 251101 (2010),
doi:https://doi.org/10.1103/PhysRevLett.104.251101.
\bibitem{Caballero2012AJ} O. L. Caballero, G. C. McLaughlin and R. Surman,\emph{ Astrophys. J.}
\textbf{745} (2012) 170, doi:10.1088/0004-637X/745/2/170.
\bibitem{Garvey1969RMP}G. T. Garvey, W. J. Gerace, R. L. Jaffe, I. Talmi and I.
Kelson, \emph{Rev. Mod. Phys.} \textbf{41} (1969) S1,
doi:https://doi.org/10.1103/RevModPhys.41.S1.
\bibitem{Zhang1989PLB}J. Y. Zhang, R. F. Casten and D. S. Brenner, \emph{Phys. Lett. B}
\textbf{227} (1989) 1,
doi:https://doi.org/10.1016/0370-2693(89)91273-2.
\bibitem{Fu2010PRC}G. J. Fu, H. Jiang, Y. M. Zhao, S. Pittel and A. Arima, \emph{Phys.
Rev. C} \textbf{82} (2010) 034304,
doi:https://doi.org/10.1103/PhysRevC.82.034304.
\bibitem{Sun2011SCPMA}B. Sun, P. Zhao and J. Meng, \emph{Sci. China Phys. Mech. Astron.}
\textbf{54} (2011) 210, doi:10.1007/s11433-010-4222-8.
\bibitem{Kaneko2013PRL}K. Kaneko, Y. Sun, T. Mizusaki and S. Tazaki, \emph{Phys. Rev. Lett.}
\textbf{110} (2013) 172505,
doi:https://doi.org/10.1103/PhysRevLett.110.172505.
\bibitem{Dong2011PRL}J. M. Dong, W. Zuo and W. Scheid, \emph{Phys. Rev. Lett.} \textbf{107} (2011)
012501, doi:https://doi.org/10.1103/PhysRevLett.107.012501.
\bibitem{Mo^{"}ller1995ADNDT}P. M\"{o}ller, J. R. Nix, W. D. Myers and W. J. Swiatecki, \emph{At. Data Nucl.
Data Tables }\textbf{59} (1995) 185,
doi:http://dx.doi.org/10.1006/adnd.1995.1002.
\bibitem{Wang2014PLB}N. Wang, M. Liu, X. Z. Wu, J. Meng, \emph{Phys. Lett. B} \textbf{734} (2014)
215, doi:http://dx.doi.org/10.1016/j.physletb.2014.05.049.
\bibitem{Goriely2013PRC}S. Goriely, N. Chamel and J. M. Pearson, \emph{Phys. Rev. C} \textbf{88} (2013)
061302(R), doi:https://doi.org/10.1103/PhysRevC.88.061302.
\bibitem{Geng2005PTP}L. S. Geng, H. Toki and J. Meng, \emph{Prog. Theor. Phys.} \textbf{113} (2005)
785, doi:https://doi.org/10.1143/PTP.113.785.
\bibitem{Hua2012SCPMA}X. M. Hua, T. H. Heng, Z. M. Niu, B. H. Sun and J. Y. Guo, \emph{Sci. China: Phys. Mech. Astron.} \textbf{55} (2012)
2414, doi: 10.1007/s11433-012-4943-y.
\bibitem{Arteaga2016EPJA}D. Pe\~{n}a-Arteaga, S. Goriely and N. Chamel, \emph{Eur. Phys. J. A} \textbf{52} (2016)
320, doi:10.1140/epja/i2016-16320-x.
\bibitem{Morales2010PRC}Irving O. Morales, P. Van Isacker, V. Velazquez, J. Barea, J. Mendoza-Temis, J. C. L\'{o}pez Vieyra, J. G. Hirsch and A. Frank,
\emph{Phys. Rev. C} \textbf{81} (2010) 024304,
doi:https://doi.org/10.1103/PhysRevC.81.024304.
\bibitem{Wang2011PRC}N. Wang, M. Liu, \emph{Phys. Rev. C} \textbf{84} (2011)
051303(R), doi:https://doi.org/10.1103/PhysRevC.84.051303.
\bibitem{Niu2013PRCb}Z. M. Niu, Z. L. Zhu, Y. F. Niu, B. H. Sun, T. H. Heng and J. Y. Guo, \emph{Phys. Rev. C} \textbf{88} (2013)
024325, doi:https://doi.org/10.1103/PhysRevC.88.024325.
\bibitem{Zheng2014PRC}J. S. Zheng, N. Y. Wang, Z. Y. Wang, Z. M. Niu, Y. F. Niu and B. Sun, \emph{Phys. Rev. C} \textbf{90} (2014)
014303, doi:https://doi.org/10.1103/PhysRevC.90.014303.
\bibitem{Niu2016PRC}Z. M. Niu, B. H. Sun, H. Z. Liang, Y. F. Niu and J. Y. Guo, \emph{Phys. Rev. C} \textbf{94} (2016)
054315, doi:https://doi.org/10.1103/PhysRevC.94.054315.
\bibitem{Niu2018SciB}Z. M. Niu, H. Z. Liang, B. H. Sun, Y. F. Niu, J. Y. Guo and J. Meng, \emph{Sci. Bull.} \textbf{63} (2018)
759, doi:https://doi.org/10.1016/j.scib.2018.05.009.
\bibitem{Utama2016PRC}R. Utama, J. Piekarewicz and H. B. Prosper, \emph{Phys. Rev. C} \textbf{93} (2016)
014311, doi:https://doi.org/10.1103/PhysRevC.93.014311.
\bibitem{Zhang2017JPG}H. F. Zhang, L. H. Wang, J. P. Yin, P. H. Chen and H. F. Zhang, \emph{J. Phys. G: Nucl. Part.
Phys.} \textbf{44} (2017) 045110,
doi:https://doi.org/10.1088/1361-6471/aa5d78.
\bibitem{Niu2018PLB}Z. M. Niu and H. Z. Liang, \emph{Phys. Lett. B} \textbf{778} (2018)
48, doi:https://doi.org/10.1016/j.physletb.2018.01.002.
\bibitem{Burbidge1957RPM}E. M. Burbidge, G. R. Burbidge, W. A. Fowler and F. Hoyle, \emph{Rev. Mod. Phys.}
\textbf{29} (1957) 547,
doi:https://doi.org/10.1103/RevModPhys.29.547.
\bibitem{Hilf1976CERN}E. R. Hilf, H. V. Groote and K. Takahashi, in Proceedings of the
Third International Conference on Nuclei Far from Stability (CERN,
Geneva, 1976), p. 142.
\bibitem{Pfeiffer1997ZPA}B. Pfeiffer, K. -L. Kratz and F. -K. Thielemann, \emph{Z. Phys. A} \textbf{357} (1997)
235, doi:10.1007/s002180050237.
\bibitem{Wanajo2004AJ}S. Wanajo, S. Goriely, M. Samyn and N. Itoh, \emph{Astrophys. J.} \textbf{606} (2004) 1057,
doi:http://dx.doi.org/10.1086/383140.
\bibitem{Sun2008PRC}B. Sun, F. Montes, L. S. Geng, H. Geissel, Yu. A. Litvinov and J.
Meng, \emph{Phys. Rev. C} \textbf{78} (2008) 025806,
doi:https://doi.org/10.1103/PhysRevC.78.025806.
\bibitem{Niu2009PRC}Z. M. Niu, B. Sun and J. Meng, \emph{Phys. Rev. C} \textbf{80} (2009)
065806, doi:https://doi.org/10.1103/PhysRevC.80.065806.
\bibitem{Xu2013PRC}X. D. Xu, B. Sun, Z. M. Niu, Z. Li, Y. Z. Qian and J. Meng,
\emph{Phys. Rev. C} \textbf{87} (2013) 015805,
doi:https://doi.org/10.1103/PhysRevC.87.015805.
\bibitem{Fermi1934ZP}E. Fermi, \emph{Z. Phys. A} \textbf{88} (1934) 161,
doi:10.1007/BF01351864.
\bibitem{Engel1999PRC}J. Engel, M. Bender, J. Dobaczewski, W. Nazarewicz and R. Surman, \emph{Phys. Rev. C} \textbf{60} (1999)
014302, doi:https://doi.org/10.1103/PhysRevC.60.014302.
\bibitem{Liang2008PRL}H. Z. Liang, N. Van Giai and J. Meng, \emph{Phys. Rev. Lett.} \textbf{101} (2008)
122502, doi:https://doi.org/10.1103/PhysRevLett.101.122502.
\bibitem{Niu2017PRC}Z. M. Niu, Y. F. Niu, H. Z. Liang, W. H. Long and J. Meng, \emph{Phys. Rev. C} \textbf{95} (2017)
044301, doi:https://doi.org/10.1103/PhysRevC.95.044301.
\bibitem{Moller1997ADNDTMoller2003PRC}P. M\"{o}ller, J. R. Nix and K.-L. Kratz, \emph{At. Data Nucl. Data Tables} \textbf{66} (1997)
131, doi:1-s2.0-S0092640X97907464; P. M\"{o}ller, B. Pfeiffer and
K.-L. Kratz, \emph{Phys. Rev. C} \textbf{67} (2003) 055802,
doi:https://doi.org/10.1103/PhysRevC.67.055802.
\bibitem{Borzov2000PRC}I. N. Borzov and S. Goriely, \emph{Phys. Rev. C} \textbf{62} (2000)
035501, doi:https://doi.org/10.1103/PhysRevC.62.035501.
\bibitem{Borzov1996ZPA}I. N. Borzov, et al., \emph{Z. Phys. A} \textbf{355} (1996)
117, doi:10.1007/s002180050088.
\bibitem{Niu2013PLB}Z. M. Niu, Y. F. Niu, H. Z. Liang, W. H. Long, T. Nik\v{s}i\'{c}, D. Vretenar and J. Meng, \emph{Phys. Lett. B }\textbf{723} (2013)
172, doi:http://dx.doi.org/10.1016/j.physletb.2013.04.048.
\bibitem{Niu2013PRCR}Z. M. Niu, Y. F. Niu, Q. Liu, H. Z. Liang and J. Y. Guo, \emph{Phys. Rev. C}
\textbf{87} (2013) 051303(R),
doi:https://doi.org/10.1103/PhysRevC.87.051303.
\bibitem{Marketin2016PRC}T. Marketin, L. Huther and G. Mart¨ªnez-Pinedo, \emph{Phys. Rev. C} \textbf{93} (2016)
025805, doi:https://doi.org/10.1103/PhysRevC.93.025805.
\bibitem{Zhang2006PRCZhang2007JPGNPP} X. Zhang and Z. Ren, \emph{Phys. Rev. C} \textbf{73} (2006)
014305, doi:https://doi.org/10.1103/PhysRevC.73.014305; X. Zhang, Z.
Ren, Q. Zhi and Q. Zheng, \emph{J. Phys. G-Nucl. Part. Phys.}
\textbf{34} (2007) 2611,
doi:https://doi.org/10.1088/0954-3899/34/12/007.
\bibitem{Zhou2017SCPMA}Y. Zhou, Z. H. Li, Y. B. Wang, Y. S. Chen, B. Guo, J. Su, Y. J. Li,
S. Q. Yan, X. Y. Li, Z. Y. Han, Y. P. Shen, L. Gan, S. Zeng, G. Lian
and W. P. Liu, \emph{Sci. China-Phys. Mech. Astron.} \textbf{60}
(2017) 082012, doi: 10.1007/s11433-017-9045-0.
\bibitem{Audi2017CPC} G. Audi, F. G. Kondev, M. Wang, W. J. Huang and S. Naimi, \emph{Chin. Phys. C} \textbf{41(3)} (2017)
30001-030001, doi:10.1088/1674-1137/41/3/030001.
\bibitem{Mathews1990N} G. J. Mathews and J. J. Cowan, Nature (London) \textbf{345} (1990)
491, doi:https://doi.org/10.1038/345491a0.
\bibitem{Kratz2007AJ} K.-L. Kratz, K. Farouqi, B. Pfeiffer, J. W. Truran, C. Sneden
and J. J. Cowan, \emph{Astrophys. J.} \textbf{662} (2007)
39,doi:http://dx.doi.org/10.1086/517495.
\bibitem{Cowan1999AJ} J. J. Cowan, B. Pfeiffer, K. -L. Kratz, F. -K. Thielemann, C.
Sneden, S. Burles, D. Tyler and T. C. Berrs,\emph{ Astrophys. J.}
\textbf{521} (1999) 194, doi:http://dx.doi.org/10.1086/307512.
\bibitem{Simmerer2004AJ}J. Simmerer, C. Sneden, J. J. Cowan, J. Collier, V. M. Woolf and J. E. Lawler, \emph{Astrophys. J.} \textbf{617} (2004) 1091,
doi:http://dx.doi.org/10.1086/424504.
\bibitem{Goriely2016PRC}S. Goriely, N. Chamel and J. M. Pearson, \emph{Phys. Rev. C} \textbf{93} (2016)
034337, doi:https://doi.org/10.1103/PhysRevC.93.034337.
\bibitem{Wang2017CPC} M. Wang, G. Audi, F. G. Kondev, W. J. Huang, S. Naimi and Xing Xu, \emph{Chin. Phys. C} \textbf{41(3)} (2017)
30003-030003, doi:10.1088/1674-1137/41/3/030003.


\end{thebibliography}
\end{document}